\documentclass[epjCONF,columns]{svjour} 
\usepackage{graphicx}
\usepackage[varg]{txfonts} 
\usepackage[latin1]{inputenc}
\session-title{Conference Title, to be filled}
\begin{document}
\title{Effective Field Theory for Top and Weak Boson Physics}
\author{Scott Willenbrock\thanks{\email{willen@illinois.edu}}
}
\institute{Department of Physics, University of Illinois at Urbana-Champaign \\
1110 West Green Street, Urbana, IL  61801}
\abstract{
Effective field theory is the ideal framework for discussing top and weak boson properties.
We discuss the application of this framework to top physics at both tree level and one loop.
We consider weak boson pair production within an effective field theory framework, and argue that one need not be concerned with the violation of unitarity bounds at energies beyond the region where there are data.
} 
\maketitle
\section{Introduction}

Why study top and weak boson properties?  There are two reasons:
\begin{itemize}
\item To measure fundamental parameters of the Standard Model,
such as $m_t$ and $M_W$.
\item To search for physics beyond the Standard Model.
\end{itemize}

\begin{figure}[htb]
\centering
\resizebox{.8\columnwidth}{!}{  \includegraphics
{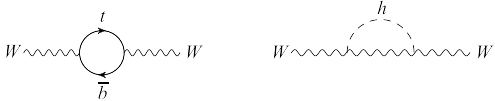} }
\caption{The top quark and the Higgs boson contribute to precision electroweak measurements
at one loop.}
\label{fig:1}       
\end{figure}

\begin{figure}[hbt]
\centering
\resizebox{\columnwidth}{!}{
\includegraphics
{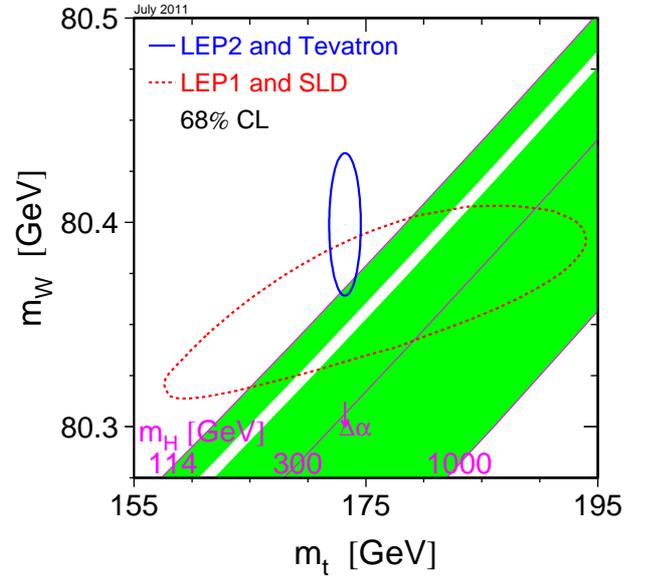}
}
\caption{The $W$ boson mass {\it vs.} the top quark mass.  The diagonal lines correspond to a given Higgs mass.  The direct measurements, indicated by the narrow vertical (blue) ellipse, indicate a Higgs boson mass near the current lower bound of 114 GeV.}
\label{fig:2}       
\end{figure}

Let's begin with the first motivation. Actually, neither $m_t$ nor $M_W$
are truly fundamental parameters, in that they are both derived from
more fundamental quantities: $m_t=y_tv/\sqrt 2$ and $M_W=gv/2$, where
$v$ is the Higgs vacuum expectation value (known from the Fermi
constant, $G_F=1/\sqrt 2v^2$).  Thus when we measure $m_t$ we are really
measuring the Yukawa coupling $y_t$.  When we measure $M_W$ we are
measuring the weak coupling $g$, which we already know more accurately
from other measurements.  Our real interest in measuring $M_W$ is that
it is sensitive, via the loop diagrams in Fig.~\ref{fig:1}, to the top and Higgs
masses.  Since we know the top mass quite accurately now, a precise
measurement of $M_W$ indicates the preferred mass for the Higgs boson, as
shown in Fig.~\ref{fig:2}. The data prefer a relatively light Higgs boson, near the current lower bound of 114 GeV.

That brings us to our second motivation.  We look for physics beyond the
Standard Model in top and weak boson properties by looking for
deviations from the Standard Model.  This raises the question of the
best way to parameterize such deviations.  I will argue that the answer
is effective field theory \cite{Weinberg:1978kz}.  Before I do, let me remind you about
effective field theory via an example.

Consider the Standard Model with a new particle introduced, a $Z^\prime$
boson that couples to ordinary fermions.  At energies above the
$Z^\prime$ mass, one can just produce it directly.  At energies
below the $Z^\prime$ mass, one can only see its indirect effects via the
exchange of a virtual $Z^\prime$ between ordinary fermions.  At low
energy this looks like a new four-fermion interaction,
\begin{equation}
{\cal L} = {\cal L}_{SM} + \frac{g^{\prime 2}}{M^2_{Z^{\prime}}}\bar\psi\psi\bar\psi\psi
\end{equation}
where $g^{\prime}$ is the coupling of
the $Z^\prime$ to ordinary fermions.  This four-fermion operator is of
dimension six, because fermion fields are of dimension 3/2.  This is in contrast to the Standard Model interactions, which are
of dimension four or less.

In an effective field theory, we abstract these ideas and write
\begin{equation}
{\cal L} = {\cal L}_{SM} + \sum_i \frac{C_i}{\Lambda^2}{\cal O}_i+\cdots
\end{equation}
where ${\cal O}_i$ are dimension six operators constructed from Standard
Model fields.  The new physics resides at energy $\Lambda$, which is
greater than the experimentally accessible energies, and the
coefficients $C_i$ are a measure of the strength with which the new
physics couples to ordinary particles.

The bad news is that there are 59 independent dimension six operators,
even for just one generation of quarks and leptons \cite{Buchmuller:1985jz,Grzadkowski:2010es}.  The good news is
that only a few of these operators are pertinent to top and weak boson
physics.

Let's consider a simple example, top quark decay to zero-helicity $W$
bosons.  The branching ratio at tree level is
$m_t^2/(m_t^2+M_W^2)\approx 0.7$.  If we neglect the $b$ quark mass,
there is only one dimension six operator that contributes at leading
order in $1/\Lambda^2$, namely ${\cal O}_{tW} = (\bar
q\sigma^{\mu\nu}\tau^It)\tilde\phi W^I_{\mu\nu}$ (where $q$ is the left-chiral ($t,b$) doublet, $t$ is the right-chiral top, and $\phi$ is the Higgs doublet).  This operator
modifies the helicity-zero branching ratio to \cite{AguilarSaavedra:2006fy,Zhang:2010dr}
\begin{equation}
f_0 = \frac{m_t^2}{m_t^2+M_W^2}-\frac{4\sqrt 2C_{tW}
v^2}{\Lambda^2}\frac{m_tM_W(m_t^2-M_W^2)}{(m_t^2+2M_W^2)^2}
\end{equation}
By comparing with the measured value of $f_0$, one can extract a bound
on the coefficient of the dimension six operator: $C_{tW}/\Lambda^2=0.03\pm 0.94$
TeV$^{-2}$.

There are a set of five dimension six operators, including ${\cal
O}_{tW}$, that can be measured or bounded from top decay \cite{AguilarSaavedra:2006fy,Zhang:2010dr}, single top
production \cite{Zhang:2010dr,Cao:2007ea,AguilarSaavedra:2008gt}, and $t\bar t$ production \cite{Zhang:2010dr,Cho:1994yu,Degrande:2010kt} at the LHC.  A strategy for making
these measurements is outlined in Ref.~\cite{Zhang:2010px}.  Additional operators may be included in the analyses if desired \cite{AguilarSaavedra:2010sq}.

We can also bound the coefficient of dimension six operators involving
the top quark by looking at their contribution to precision electroweak
measurements at one loop.  As we mentioned earlier in Fig.~1, the top
quark makes a contribution to the $W$ boson mass at one loop.  If we
replace one of the vertices in the loop diagram with a vertex arising
from a dimension six operator, then we may be able to place a bound on
the coefficient of this operator.  For example, the operator ${\cal
O}_{tW}$ discussed above contributes to the oblique parameter $\hat U$
\cite{Barbieri:2004qk} an amount \cite{Greiner:2011tt}
\begin{equation}
\hat U = N_c \frac{gC_{tW}}{4\pi^2}\frac{\sqrt 2vm_t}{4\Lambda^2}
\end{equation}
where $N_c=3$ is the number of colors.  The value of the $\hat U$
parameter is dominated by the $W$ boson mass.  Comparing with data yields
the bound $C_{tW}/\Lambda^2=-0.7\pm 1.1$ TeV$^{-2}$, comparable in precision to
the bound from top decay to helicity-zero $W$ bosons given above.  One can perform
a similar one-loop analysis for other dimension six operators involving the top
quark \cite{Zhang:2012cd}.

Let's pause and consider the many virtues of the effective field theory
approach to physics beyond the Standard Model:
\begin{itemize}
\item It is well motivated, and provides guidance as to the most likely
places to observe the effects of new physics.
\item It respects the known $SU(3)_C\times SU(2)_L\times U(1)_Y$ gauge
symmetries of nature.
\item It includes not only nonstandard vertex interactions, but also self energy and
contact interactions, such as a $\bar du\bar tb$ interaction that
affects single-top production.
\item It is valid whether the particles involved are real or virtual.
\item It allows one to calculate at both tree and loop level, as
evidenced by the calculation of the $\hat U$ parameter mentioned above.
\end{itemize}
No other approach to physics beyond the Standard Model is as robust. However,
it must be kept in mind that an effective
field theory is a low-energy theory, valid only up to the scale of
new physics, $\Lambda$.

\begin{figure}[hbt]
\centering
\resizebox{.5\columnwidth}{!}{  \includegraphics
{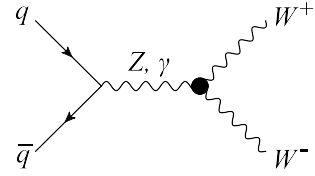} }
\caption{Weak boson pair production is sensitive to dimension six operators that modify the weak boson self interactions.}
\label{fig:3}       
\end{figure}

Let's now turn to an effective field theory approach to weak boson
physics, in particular to weak boson pair production, shown in Fig.~\ref{fig:3}.
The coupling of the weak bosons to fermions is much better constrained
than the weak boson self interactions, so we will only consider
dimension six operators that contribute to the latter.  There are five
independent operators: three conserve $C$ and $P$ \cite{Hagiwara:1993ck},
\begin{eqnarray}
{\cal O}_{WWW}&=&\mbox{Tr}[W_{\mu\nu}W^{\nu\rho}W_{\rho}^{\mu}]\\
{\cal O}_W&=&(D_\mu\Phi)^\dagger W^{\mu\nu}(D_\nu\Phi)\\
{\cal O}_B&=&(D_\mu\Phi)^\dagger B^{\mu\nu}(D_\nu\Phi)
\end{eqnarray}
and two violate $C$ and/or $P$:
\begin{eqnarray}
{\cal O}_{\tilde WWW}&=&\mbox{Tr}[{\tilde
W}_{\mu\nu}W^{\nu\rho}W_{\rho}^{\mu}]\\
{\cal O}_{\tilde W}&=&(D_\mu\Phi)^\dagger {\tilde W}^{\mu\nu}(D_\nu\Phi)
\end{eqnarray}
There is no reason to exclude the $C$ and/or $P$ violating operators
unless the physics beyond the Standard Model respects these discrete
symmetries.

\begin{figure}[hbt]
\centering
\resizebox{\columnwidth}{!}{  \includegraphics
{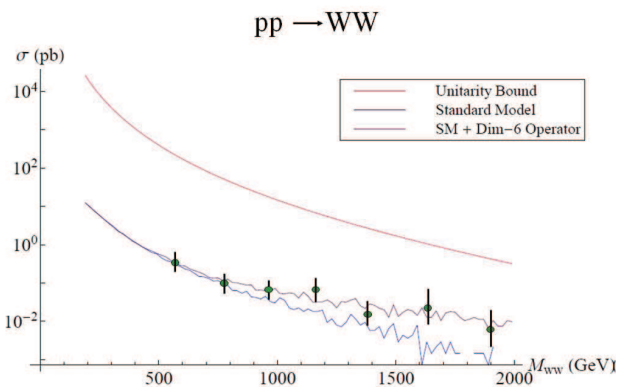} }
\caption{The invariant mass distribution for $W^+W^-$ production at the LHC.
The lowest (blue) curve is the Standard Model prediction; the middle (purple) curve includes the effect of the dimension six operator ${\cal O}_{WWW}$; and the upper (red) curve is the
unitarity bound.  The data are hypothetical.}
\label{fig:4}       
\end{figure}

If present, these operators will cause the cross section for weak boson
pair production to deviate from the Standard Model prediction at high
invariant mass.  An example for $W^+W^-$ production at the LHC is shown
in Fig.~\ref{fig:4}.  The lowest (blue) curve is the Standard Model prediction,
superimposed on hypothetical data points which deviate from the
prediction.  The middle (purple) curve includes the effects of the dimension six
operator ${\cal O}_{WWW}$, with the coefficient adjusted to give a good
fit to the data.

Traditionally, weak boson pair production has been considered in the
framework of anomalous couplings rather than effective field theory \cite{Gaemers:1978hg,Hagiwara:1986vm}.
However, we can relate the two approaches.  Let's restrict our attention
to the three $C$ and $P$ conserving operators above.  From them, one can
derive the five $C$ and $P$ conserving anomalous couplings \cite{Hagiwara:1993ck}:
\begin{eqnarray}
g_1^Z & = & 1+c_W\frac{m_Z^2}{2\Lambda^2}\\
\kappa_\gamma & = & 1+(c_W+c_B)\frac{m_W^2}{2\Lambda^2}\\
\kappa_Z & = & 1+(c_W-c_B\tan^2\theta_W)\frac{m_W^2}{2\Lambda^2}\\
\lambda_\gamma & = & \lambda_Z = c_{WWW}\frac{3g^2m_W^2}{2\Lambda^2}
\end{eqnarray}
The effective field theory approach is simpler, in that there are just three independent parameters instead of five.  Furthermore, the effective field theory approach clarifies that the anomalous couplings are truly constants, independent of energy.  This follows from the fact that the coefficients of the dimension six operators are constants.

This last point deserves some discussion, because anomalous couplings have often been taken to be form factors, with the rationale that they must be suppressed at high energy in order for the cross section to respect the unitarity bound \cite{Zeppenfeld:1987ip}. Let's take a fresh look at this argument, from the perspective of an effective field theory.

A simple argument shows that we need not be concerned with unitarity bounds at all.  The data necessarily respect the unitarity bound, because the bound is a physical requirement.  A theory that fits the data will therefore also respect the bound, at least in the region where there is data.  This is exemplified in Fig.~4.  The upper (red) curve is the unitarity bound, and since the data cannot exceed this bound, the theoretical fit to the data also cannot exceed the bound, at least in the region where there is data.

If we extend the theoretical curve to higher energy, beyond the region where there is data, it will eventually violate the unitarity bound.  But recall that the theoretical curve is generated by an effective field theory, which is valid only up to the scale of new physics, $\Lambda$.  An effective field theory does not have the ambition to describe nature to arbitrarily high energies, only to fit the data.

We have argued that effective field theory is the ideal way to describe top and weak boson properties.  Along with its many virtues, listed above, we have argued that unitarity bounds on physical cross sections are automatically satisfied by an effective field theory.  There is no need for the awkward and arbitrary introduction of form factors in weak boson pair production.  More details may be found in Ref.~\cite{degrand}.
%

%
%

\end{document}